\documentclass[preprint,12pt]{revtex4}

\usepackage{amssymb}
\usepackage{graphicx}
\begin{document}
	
%	\begin{frontmatter}
		
		
\title{Complexity in cancer stem cells and tumor evolution: towards precision medicine}  %$^\dag$}}

\author{Caterina A. M. La Porta $^{\ast}$\textit{$^{a,b}$}, Stefano Zapperi\textit{$^{a,c,d,e,f}$} }
. 
\address{$^{a}$ Center for Complexity and Biosystems, 
			University of Milan, via Celoria 16, 20133 Milano, Italy.\\
			$^{b}$ Department of Biosciences, University of Milan, via Celoria 26, 20133 Milano, Italy\\
	$^{c}$ Department of Physics, University of Milan, via Celoria 16, 20133 Milano, Italy. \\
	$^{d}$ Institute for Scientific Interchange Foundation,  Via Alassio 11/C, 10126 Torino, Italy\\
	$^{e}$ Department of Applied Physics, Aalto University, P.O. Box 11100, FIN-00076 Aalto, Espoo, Finland\\
	$^{f}$ CNR - Consiglio Nazionale delle Ricerche,  ICMATE, Via R. Cozzi 53, 20125 Milano, Italy\\
	$^{\ast}$ Corresponding author: caterina.laporta@unimi.it}

\begin{abstract}
In this review, we discuss recent advances on the plasticity of cancer stem cells 
and highlight their relevance to understand the metastatic process and to guide therapeutic
interventions. Recent results suggest that the strict hierarchical structure of cancer cell populations
advocated by the cancer stem cell model must be reconsidered since the depletion of cancer stem
cells leads the other tumor cells to switch back into the cancer stem cell phenotype. This plasticity
has important implications for metastasis since migrating cells do not need to be cancer stem cells 
in order to seed a metastasis. We also discuss the important role of the immune system and the microenvironment in 
modulating phenotypic switching and suggest possible avenues to
exploit our understanding of this process to develop an effective strategy for precision medicine.
\end{abstract}

%\begin{keyword}
%	\sep cancer stem cells \sep phenotypic switching \sep metastasis \sep precision medicine
	
	
%\end{keyword}
\maketitle

%\end{frontmatter}

\section*{Do tumors grow stochastically or hierarchically?}
Two main models have been used in the past to describe tumor development: in the stochastic
clonal evolution model \citep{Nowell1976}, each cells can be tumorigenic and sustain tumor growth, while according to the cancer stem cell (CSC) theory \citep{bonnet1997}, the tumor is organized hierarchically with a subpopulation of cells sustaining the growth. The relevance of either one or the other theory has a relevance for therapeutic strategies \citep{Jordan2006,Clevers2011}. Whereas the stochastic model suggests that the best therapeutic strategy is to kill all cancer cells targeting some common factors, according to the CSC theory the best strategy should be to  primarily target CSCs inside the heterogeneous population. The main idea coming from this debate is that once we answer the question of how   tumors evolve, either stochastically or hierarchically, we could tackle the issue of the impact of each theory on metastasis. There are papers providing strong evidence of the existence of a CSCs population in different tumors,
including observations in vivo \cite{snippert2010}, but contrasting results appeared regularly in the literature.
A possible explanation for this controversy might come from plasticity: the possibility that
cancer cells might revert to the CSC state \cite{gupta2011,Auffinger2014,Iliopoulos2011,Vlashi2015,Ohanna2013,sellerio2015}.  In the following we will discuss this issue, focusing on the relevance to metastasis.

\begin{figure}[hb]
	\centering
	\includegraphics[width=15cm]{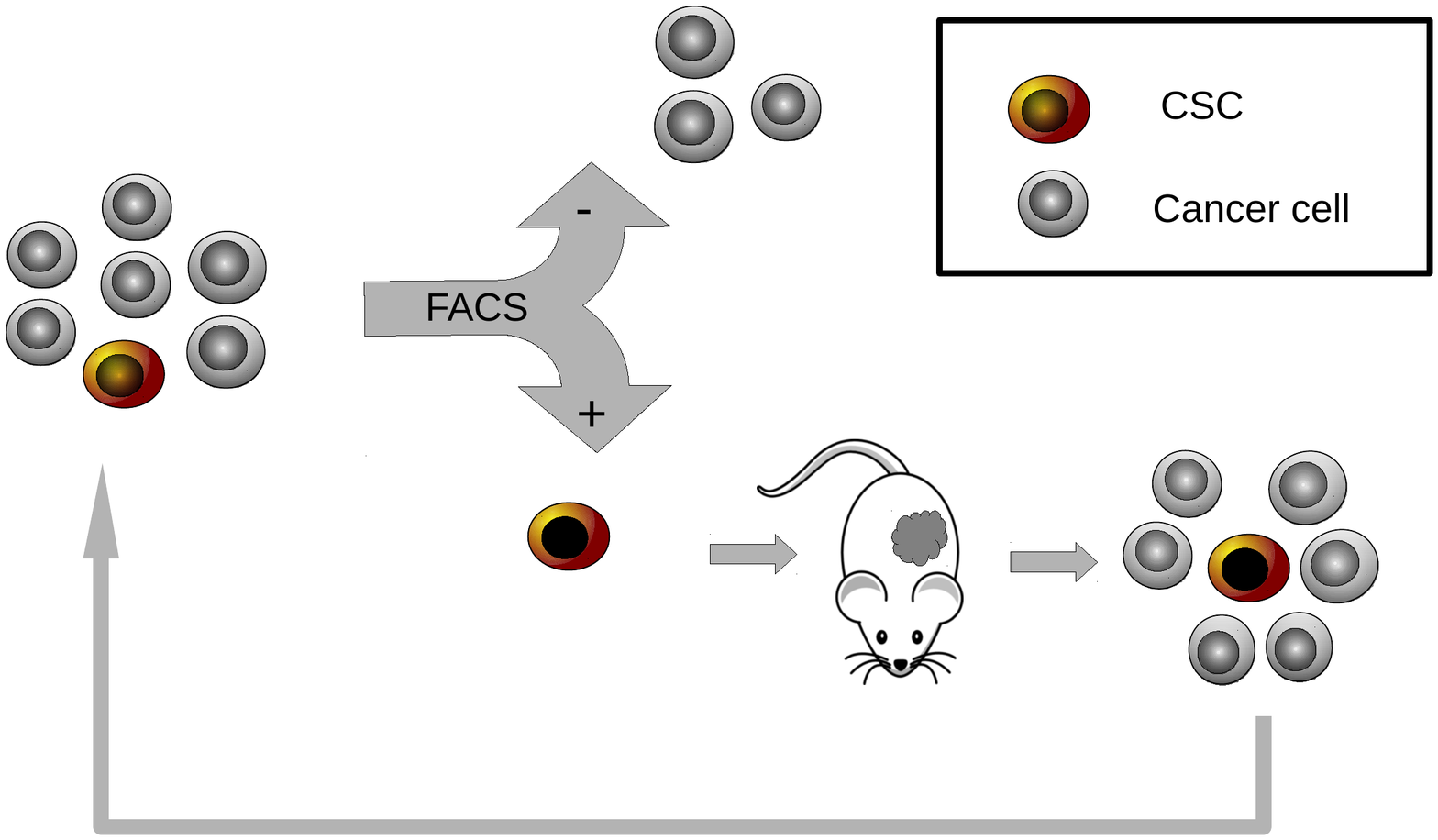}
	\caption{{\bf Identification of CSCs.} CSCs are identified by FACS sorting followed by transplantation in mice. The process is then repeated.
		\label{fig:fig1}}
\end{figure}

The CSC was defined as a cell within the tumor that has the capacity of self-renewing and generating an heterogeneous population of cancer cells composing the tumor \citep{Jordan2006,Clevers2011}. Contrary to normal stem cells, however, where proliferation capabilities and genomic integrity are tightly regulated and controlled, in CSCs these controls are typically missing.  In practice, CSCs can only be defined experimentally by their ability to recapitulate the generation of a continuously growing tumor. To this end, putative CSCs were identified according to the expression of surface markers (e.g factors expressed by normal stem cells) and isolated through fluorescence-activated cell sorting (FACS).  The cells isolated in this way are then transplanted (or engrafted) in immunocompromized mice. If the mouse develops a tumor, cancer cells expressing the CSC markers are again isolated and transplanted into mice (Fig. \ref{fig:fig1})  The observation of a series of CSC identifications and transplantations is considered an evidence for the existence of a CSC population within the tumor. Both xeno- and syngeneic transplantations might, however, misrepresent the intricate network of interactions with diverse support such as fibroblasts, endothelial cells, macrophages, mesenchymal stem cells and many of the cytokines and receptors involved in these interactions \citep{laporta2009}. Another important problem is that there is no clear recipe to choose the best marker to unambiguously identify CSCs. One is thus forced to proceed by trial and error, exploiting analogies with normal stem cells. 

Taking into account all these limitations, the first evidence of CSCs came from hematological tumors \citep{bonnet1997} and later from solid tumors such as breast \cite{Al-Hajj2003,Park2010,Charafe-Jauffret2009}, prostate \citep{Tang2007}, brain cancer \citep{Salmaggi2006} and melanoma \citep{Fang2005,hadnagy2006,Dou2007,monzani2007,schatton2008,klein2007}. In breast cancer, the first evidence of a subpopulation with a specific cell-surface antigen profile (CD44+/CD24-) that successfully establish itself as tumor  xenograft was published in 2003 
\cite{Al-Hajj2003}. In another study, 275 patients were analyzed for CD44+/CD24- putative stem cell marker as well as for others (vimentin, ostenectin, connexin 43, ADLH, CK18, GATA3, MUC1) in primary breast cancers of different subtypes and hystological stages. This study reveals a high degree of diversity in the expression of several of the selected markers in different tumor subtypes and hystologic stages \citep{Park2010}. Furthermore, aldehyde dehydrogenase (ALDH) was used as stem cell marker in 33 human breast cell lines \citep{Charafe-Jauffret2009}. ALDH is a detoxifying enzyme that oxidizes intracellular aldehydes and it is thought to play a role in the differentiation  of stem cells via the metabolism of retinal to retinoic acid \citep{Chute2006}.  Interestingly, ALDH activity can be used to sort a subpopulation of cells that display stem cell properties from normal breast tissue and breast cancer \citep{Ginestier2007} and to isolate CSCs from multiple myeloma and acute leukemia as well as from brain tumors \citep{Cheung2007,Corti2006}. The ALDH phenotype was not associated with more-aggressive subpopulations in melanoma, suggesting that it is not a "universal" marker \citep{Prasmickaite2010}.  Several markers that select aggressive subpopulations have been identified in glioblastoma multiforme \citep{Salmaggi2006}.  Similarly, several candidate populations of prostate stem/progenitor cells have been reported including those expressing high levels of CD44, integrin $\alpha$2$\beta$1, or  CD133 \citep{Tang2007}.  Interestingly, two independent studies in the mouse prostate have identified two different stem cell populations. One, marked by CD117 (c-Kit), seems to be localized in the basal layer \citep{Leong2008} and and the other, called castration-resistant Nkx3.1-expressing cells, in the luminal layer \citep{Wang2009}. Identification and characterization of normal prostate stem cells is clearly relevant to understand the origin for human prostatic cancer \citep{Li2011}. This is because it is  difficult to ascertain the potential overlap as well as lineage relationships of the various candidate stem cells that have been identified \citep{Shen2010}.  This in part is due to the distinct methodologies and assays employed \citep{Shen2010}. 

Considering melanoma, several papers provided evidence supporting the existence of a CSCs subpopulation \citep{Fang2005,hadnagy2006,Dou2007,monzani2007,schatton2008,klein2007}. In 2008, however, a paper argued against CSCs based on the following observations: a relatively large fraction of melanoma cells (up to ~25\%) was shown to initiate tumors in severely immunocompromised NOD/SCID IL2R$\gamma$null mice; the fraction of tumor-inducing cells depended upon assay conditions; several putative CSC markers appeared to be reversibly expressed \citep{quintana2008}. The paper suggested that the best experimental model to confirm the presence of CSCs is severe immunocompromised mice. The authors analyzed the expression of more than 50 surface markers on melanoma cells derived from several patients (A2B5, cKIT, CD44, CD49B, CD49D, CD49F, CD133, CD166) but then mostly focused on CD133 and CD166 \citep{quintana2008}.  In a more recent paper it was shown that CD133 is highly expressed in melanoma cells and it is not a good marker to sort CSCs \citep{schatton2008}. Moreover in 2010, an independent group, using the same immunocompromised mice as in Ref. \cite{quintana2010}, did not confirm those results and proposed instead to use CD127, the nerve growth factor receptor, as a marker to identify CSCs \citep{boiko2010}. Finally, CXCR6 was proposed as a marker for a CSC-like aggressive subpopulation in human melanoma cells~\cite{taghizadeh2010}. Its peculiar feature is that it plays a critical role in stem cell biology, being linked to asymmetric cell division~\cite{taghizadeh2010}. Recent experiments in vivo have confirmed the presence of an aggressive CSC-like subpopulation in benign and malignant intestinal and skin tumors ~\citep{schepers2012,driessens2012,chen2012}. 
Contrary to previous work based on sorting and serial transplantation, these new papers tracked CSCs directly inside a growing tumor and studied the CSC population at different stages of tumor progression ~\citep{schepers2012,driessens2012,chen2012}.
Another in vitro approach that was used in recent years to investigate CSCs is to study the capability of tumor cells to form tumorospheres. The general idea is that cancer cells would not be able to form tumorspheres since they differentiate, while CSCs, resembling the characteristic of normal stem cells, can growth in a sphere. The impact of the presence of tumorspheres in physiological process like as angiogenesis and mutidrug resistance has also been studied ~\citep{Salmaggi2006}.

A marker of CSCs that is ubiquitous expressed is CD133, also known as promonin-1. This is a membrane-bound pentaspan glycoprotein that is frequentely expressed on CSC and is linked to self-renewal and tumorigenicity. CD133 is expressed in fact in hematopoietic stem cells \cite{Handgretinger2013} and was found in neural stem cells \cite{Singh2003}. It is now considered as a marker for progenitor cells\cite{Handgretinger2013,Singh2003}, but in many tumors such as melanoma it is not only restricted to the CSC subpopulation  ~\cite{monzani2007}. Therefore the advantage to use this marker is that is expressed by many CSC-tumor subpopulation but on the other side it is not always specific for CSCs.  CD133 is also related to metastasis such as in gastric cancer patients and colorectal cancer \cite{Yiming2015,Cherciu2014}
 
Despite the accumulated experimental evidence, the population dynamics of CSCs is still debated: in 2011 a paper pointed 
out the possibility that non-CSCs breast cancer cells can revert to a stem cell like state even in the absence of mutations ~\cite{gupta2011}. Similarly in melanoma, a small population of CSC-like JARID1B positive cells has been shown to be dynamically regulated in a way that differs from the standard hierarchical CSC model ~\cite{roesch2010}. This finding could reconcile the conflicting results on the existence of CSCs in melanoma ~\cite{quintana2008, quintana2010, boiko2010}. Microenvironmental factors, such as TGF$\beta$, are found to enhance the rate of switch from non-CSC cells to the CSC state ~\cite{chaffer2013}. This is in agreement with earlier results from our group showing that ABCG2 negative cells isolated from human melanoma biopsies express again this marker after few passages in vitro ~\cite{monzani2007}.  The idea that the environment is able to induce cells to switch into a more aggressive phenotype was confirmed by Medema and Vermeulen who tried to limit stemness by modifying microenvironmental factors known to support CSCs in tumors ~\cite{medema2011}. While mounting experimental evidence supports the switch from non-CSC cancer cells to the CSC state, the biological factors regulating this process are still being investigated. Two possible scenarios have been invoked to solve this puzzle: i) switch to the CSC state is either driven by genetic mutations or ii) it is regulated by epigenetic factors ~\cite{marjanovic2013}. 

Therefore the general view discussed above that only two possible scenarios are possible for tumor development, the CSC theory or the stochastic one, appears too simple.  A new concept comes into the discussion: the cellular plasticity of CSC. Many groups, including ours, proposed the concept that there is an equilibrium between CSCs and their more differentiated progeny \cite{gupta2011,Auffinger2014,Iliopoulos2011,Vlashi2015,Ohanna2013,sellerio2015}. In light of these evidence, not only the cancer cells can become CSCs but the reverse process can also happen. Our group investigated  the possible factors involved in this phenotypic switching, demonstrating that human melanoma cells can switch their phenotype to the CSC state thanks to the activation of a complex miRNA network ~\cite{sellerio2015}. In particular, 
our group showed that phenotypic switching is not a stochastic process as originally assumed~\cite{gupta2011} but it is tightly controlled by a complex network of miRNAs which in turn regulates critical pathways, such as Wnt and PI3K, leading
to phenotypic switching ~\cite{sellerio2015}. This miRNA network is activated when the CSCs are below a specific threshold implying that cancer cells are sensitive to the number of CSCs in the bulk, as also happens for normal stem cells.  Accordingly, in stem cells the locality of Wnt signaling dictates differentiation and spatial confinement in a niche ~\cite{clevers2014}.  A direct consequence of our findings is that if a reduced number of CSCs induce the other cancer cells to switch back to the CSC state. Hence,  trying to kill specifically the CSCs does not represent an effective therapeutic strategy. The capability of  cancer cells to switch to CSC under specific conditions reconcile all the contrasting reported literature. Another important consequence of these findings is that CSCs live in a niche that means a distinct microenvironment with specific functional characteristics that form the habitat of the cells. The possibility to target the niche instead of the cells could be an important way to overcome cellular resistance and prevent metastasis.

\section*{Phenotypic switching and cancer metastasis}

Metastasis represents the main problem for cancer treatment. Melanoma is a good example: most of the time 
whenever this tumor is diagnosed there is a high probability that a metastasis is already present. When a metastatic melanoma is diagnosed there are few available therapeutic strategies and their rate of success is low.  
The organs affected by metastasis depend on the probability that metastatic cancer cells reach distant organs, but also survive and grow there, initiating further metastasis. Disseminated cancer cells need supportive sites to establish a metastasis, in a way that is reminiscent of stem cell niches. For stem cells, the location and constitution of stem cell niches has been demonstrated in various tissues, including the intestinal epithelium, hematopoietic bone marrow, epidermis, and brain. In a similar way, the term metastatic niche is used to designate the specific locations, stromal cell types, visible signals, and ECM proteins that support the survival and self-renewal of disseminated metastatic cells. Metastatic cells could occupy a stem cell niche including perivascular sites recruiting stromal cells that produce stem cell niche-like components or by producing these factors themselves. Some tumors release systemic suppressor factors that make micrometastasis dormant, but others, such as melanoma, erupt decades after a primary tumor has been eradicated surgically or pharmacologically, possibly due to a reactivation of dormant micrometastasis. Recent papers show that nutrient starvation can induce autophagy causing the cancer to shrink and adopt a reversible dormancy. When the tissue changes becoming more accessible to nutrients, the tumor restarts to grow \citep{kenific2010}. Up to now, this relationship between genetic factors and microenvironment is still debated: it seems plausible that both factors contribute to metastasis.

\begin{figure}[hb]
	\centering
	\includegraphics[width=12cm]{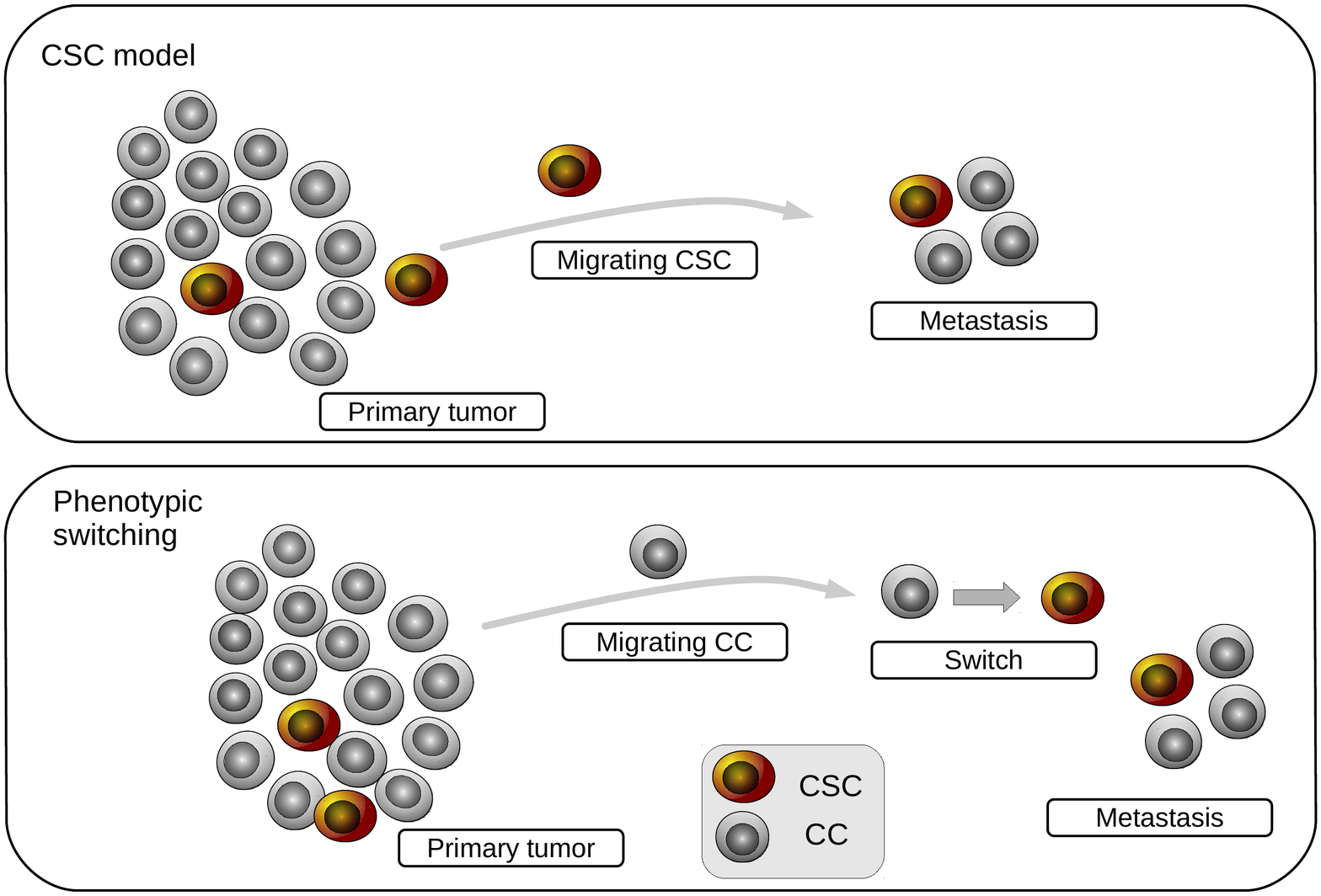}
	\caption{{\bf Phenotypic switching and metastasis.} In the CSC model, only the CSCs are able to initiate a tumor. Hence metastasis can only occur if a CSC migrates to a distant site. If phenotypic switching is possible, however, normal cancer cells (CC) can migrate and then switch into a CSC, initiating the metastatis. }\label{fig:fig2}
\end{figure}

The possibility that cancer cells can switch into CSCs under specific conditions dictated by a depletion in the number of CSCs or by the environments, opens a new 
perspective on the metastatic process. The metastatic process is usually intrepreted based on the idea that each metastasis arises from the clonal growth of a single tumor cell that has detached from tumor mass and migrated elsewhere (see Fig. \ref{fig:fig2}). According to the CSC model, a CSCs should undergo an epithelial-mesenchymal transition and then migrate to a distant site where it would seed a metastasis. The picture changes, however, if phenotypic plasticity is present: the migrating cell in this case could be a cancer cell that would then switch into a CSC at the metastatic site.
For the sake of clarity, notice that in the discussion above we mention two distinct phenotypic transformations: the epithelial-mesenchymal transition and the switching from a cancer cell to the CSC state. Throughout this review the
term "phenotypic switching" is only reserved to the second process. If phenoptypic switching is induced by the absence of CSCs in the population, as demonstraded in \cite{sellerio2015}, then a migrating cancer cell would automatically switch into a CSC once it has spread far enough from the primary tumor (see Fig. \ref{fig:fig2}). This observation is important because CSCs typically represent a small fraction of the  cancer cell population and therefore their migration would be statistically unlikely. On the other hand if  all cancer cells are potentially able to seed a metastasis by switching, the process would occur with higher probability than expected from the CSC model.

The capability to switch is driven by the microenvironment. For instance, the exposure of non stem breast cancer cells to transforming growth factor beta (TGF-$\beta$) has been shown to produce mesenchymal/CSC-like cells with a high degree of plasticity, whereas removal or inhibition of TGF-$\beta$ causes the cells to lose their mesenchymal/CSC-like characteristics, and to regain an epithelial and non-stem cell phenotype ~\cite{poleszczuk2016}.  Acquiring a mesenchymal phenotype, CSCs express characteristics that promote tumor progression recurrence, metastasis, and resistance to therapy ~\cite{liu2014b}. In a recent study, two metastatic breast cancer cell lines were exposed to repeated hypoxia/reoxygenation cycles, reproducing the typical microenvironment observed in solid tumors ~\cite{louie2010}. After one cycle of hypoxia and reoxygenation, a small subset of cells survive the hypoxia, forming spherical clusters which proliferate after reoxygenation ~\cite{louie2010}. After three of those cycles, the authors observed the appearance of a novel cell subpopulation, expressing surface marker expression (i.e., CD44++/CD24-/ESA+) associated with breast CSCs and characterized by  high tumorigenicity ~\cite{louie2010}.  Interestingly, this new subpopulation was reported to display a high fraction of CSC-like cells with respect to the original breast cancer cell lines and a stronger metastatic capacity ~\cite{louie2010}. 

The strict interplay between the plasticity of the cells and the microenvironment can thereby affect the tumor evolution and in particular the metastasis. 

\section*{Limitation of current therapies for metastasis}
CSCs are resistant to  chemotherapy and/or radiotherapy~\cite{touil2014, duru2012, phillips2006, sims2006}. The resistance might be due to multiple factors:  the tumor microenvironment which is typically rich in a diversity of proteins, including growth factors (e.g., TGF-$\beta$) and cytokines, that could activate pathways that impact the survival of CSCs ~\cite{doherty2016} and the possible role of chemo/radiotherapy in the acquisition of stemness. In fact, the problem of resistance of CSCs to conventional cancer therapies is not simply a matter of an inability of chemotherapy and radiation to destroy the CSCs but rather that the treatment itself has been shown to increase CSC characteristics in cancer cells, and can even convert non-stem cancer cells into CSCs ~\cite{wang2014, gao2016, xu2015} . Studies in breast cancer found that irradiation of breast cancer cells resulted in an increase in the number of CSCs and, in some cases, converted non tumorigenic cancer cells into CSCs ~\cite{wang2014}.  Recently, it has been found that human gastric cancer cell lines, after exposure to the chemotherapy agent 5-fluorouracil (5-FU), exhibited both resistance to 5-FU and features consistent with stemness, including tumorigenicity and capacity for self-renewal ~\cite{xu2015}. Another important example is ALDH1 which confers resistance to alkylating chemotherapeutic agents and protect against oxidative damage \cite{Huang2009,Deng2010}.  In colorectal  cancer was shown a relationship between ADLH1 expression between metastatic and non-metastaic colorectal tumors: the latter expressing high levels of ADLH1 \cite{Hessman2012}. In breast cancer a recent paper showed an increase of putative CSCs and ADHL1 positive cells after primary systemic therapy \cite{Lee2011b}. Furthermore, in head and neck cancer cells increased expression of ALDH1 led to increase expression of CSCs markers while its inhibition to a descrease of these factors \cite{Kulsum2017}. Another important factor, ABCG2, was shown to be related with chemoresistant. For example, in colorectal cancer it has been shown that the down regulation of ABCG2 inhibited the self-renewal capacity of these cells, enhancing the efficacy of chemotherapic-induced  apoptosis \cite{Ma2016}. High levels of ABCG2 was found associated to CSC-like phenotype in human melanoma \cite{monzani2007}. All these data are in agreement with the data showing that the number of CSCs are under a strict control and decreasing below a threshold increases their numbers ~\cite{sellerio2015}. This is the key factor that drives the plasticity of the tumor cells. Taking in account this new information, a successful strategy might be to impair phenotypic switching by manipulating either the cells or the microenvironment. To achieve this result, however, it is important to better understand the mechanisms that lead to phenotypic switching.  On the other hand, if we think to kill the CSCs the effect is the reverse: increase the numbers of CSCs and therefore more markers linked to a CSC-like phenotype as described above. 
On the other hand, another aspect should be consider: the resistance induced by the drugs ion the CSC's niche. In a recent interesting paper, the authors showed a chromatin-dependent mechanism exerted by let-7 which down-regulates stemness genes in cancer cells , this suggesting that epigenetic mechanisms happening in the niche are crucial for drug tolerance \cite{Liao2016}.  Similarly, in leukemia cells an alteration in the niche has been demonstrated and therefore a remodelling of the niche can be a good strategy to fight the cancer \cite{Schepers2015}. The mechanical connection with the niche of the cells has been shown also to be a crucial factor for conversion of differentiated mouse cells into a tissue-specific stem/progenitor cell state through YAP-TAZ \cite{Panciera2016}. In fact, YAP/TAZ has been shown to be able to reprogram distinct cell types to their corresponding tissue specific stem cells \cite{Panciera2016}. YAP has been shown to be regulated by the expression of cytoskeletal regulators and matrix stiffness can enhance YAP activation 
\cite{Calvo2013}. Altogether we can speculate that changes in the mechanical and chemical structure of the niche can help in reprogramming the cells and therefore in CSC plasticity.   These aspects appear crucial to develop a new therapeutic approach. It should be possible to investigate better the factors that create the niche and  how and if mechanical or chemical changes can alter it. 

\section*{CSCs, metastasis and the immune system}
It is widley accepted that the immune system is recognized and respond to tumor cells. Antigens released by the tumor cells can be passively transported in the lymph or captured and delivered by dendritic cells (DC) to regional lymph nodes via afferent lymphatic vessels. In tumor-draining lymph nodes, DCs present tumor-derived antigen in associtaion to MHC molecules and activate lymphocytes CD4+T and CD8 T effector cells  can exit the lymph node and circulate throughout the body via the bloodstream following the gradient of chemokines and adhesion molecules, extravasate and migrate into the tumor bed where they recognise the cells displaying the antigens and kill them. In principle therefore, the cancer-immunity cycle could lead to eradication of malignant cells by cytotoxic immune cells, establish tumor-specific immunological memory and prevents further tumor progression. The situation is however more complex: it could be a periferal tolerance; the migration of T cells into the tumor bed can be hindered by the disorganized vasculature and chemotactic cues ~\cite{pivarcsi2007}. The presence of inhibitor cells and molecules could impair the survival of T cells, their activation, proliferation and effector functions. A general example could be the limited therapeutic benefit of the use of antibodies against PDL1 which is often overexpressed in tumor cells, possibly due to the lack of pre-existing antitumor immunity or to the presence of other immunosuppressive mechanisms in the microenvironment. The tumor can escape from the immune surveillance using different mechanisms: from  the production of immunosuppressive molecules that attenuate the immune system, to the loss of antigen expression and/or co-stimulators, to the activation of immune suppressive factors. The possibility to use an immunotherapy strategy in an advanced and metastatic tumor is quite far to be realised. The best way forward appears 
to be the possibility to kill all the cells, both CSC and cancer cells, but in this case the toxicity could be relevant. The use of vaccines could be more interesting than the passive immunization through adoptive T cells. In fact, if the phenotypic switching mechanisms will be fully understood, it might be possible to target with a vaccine all the tumor cells in a way that does not allow them to switch. In a recent interesting paper, the authors showed that using a gene-transduced tumor cell vaccine therapy targeting CSCs they found a substantially suppression in syngenic immunocompetent mice recapitulating normal immune systems \cite{Sakamoto2017}.  This system seems very interesting to induce potent tumor-specific antitumor immunity \cite{Sakamoto2017}.  Moreover, considering the plasticity of both the tumor cells and the plasticity of immune system, the interaction between the two compartment might be protumorigenic or antitumorigenic. In another recent paper, the authors proposed a strategy to target CSCs using lysate-pulsed dendritic cells in melanoma and squamous cell carcinoma \cite{Lu2015}. The CSC-DC vaccine reduced significantly ALDH-high expressed CSCs in primary tumors and therefore seems to be useful in the adjuvant setting where local and systemic relapse are high after conventional treatment of cancers \cite{Lu2015}. The  CSC's niche as well as the interplay with the immune system might be an important filed to investigate. In this connection, an interesting theoretical paper showed that an up-regulation of glucose by tumors which can lead to active competitive resources in the tumor microenvironment between tumors and immune cells \cite{Kareva2015}. In particular, the authors proposed that CSCs can circumvent by a protective shield of cancer cells creating a barrier against the immune system and creating a competition for common resources such as glucose between the same cells and immune cells \cite{Kareva2015}. Natural killers are also interesting in view of a immune therapy. A recent paper showed that activated natural killers targeted more CSCs and therefore they might be used in immunotherapy for refractory solid malignacies suchaas metastasis \cite{Ames2015}.

\section*{Conclusion: Precision medicine and tumor plasticity}
The idea to develop a therapeutic strategy considering the heterogeneity between the patients is already mainstream in the field of tumor biology. Our new understanding of tumor plasticity and the important role of the microenvironment controlling this process provides a concrete answer to the quest for precision medicine. The discovery of phenotypic
switching into CSCs leads to challenges to existing therapeutic strategies but it also opens possible new avenues. If
cancer cells can switch back into CSCs, then targeting CSCs would be counterproductive since it would ultimately 
lead to massive replenishing of the most aggressive cancer cell population. Therefore, the key to successful strategy
would be to prevent cancer cells to switch. To this end, a possible future scenario might be to understand by microarray analysis the complex network of miRNAs produced by each tumor and study with the aid of computational analysis its possible impact on the plasticity of the cells. This could lead to identify possible critical pathways and genes to expoint for therapeutic interventions, such as the development of a correct passive or active vaccination strategies. The results reported in \cite{sellerio2015} provide a methodological guide to achieve these targets, highlighting the importance of an interdisciplinary approach.

%\bibliographystyle{nature}
%\bibliographystyle{elsarticle-num} 
%\bibliography{Book-references}

%\expandafter\ifx\csname url\endcsname\relax
%  \def\url#1{\texttt{#1}}\fi
%\expandafter\ifx\csname urlprefix\endcsname\relax\def\urlprefix{URL }\fi
%\expandafter\ifx\csname href\endcsname\relax
%  \def\href#1#2{#2} \def\path#1{#1}\fi

%\section*{Acknowledgements:}
%SZ are supported by the ERC advanced grant SIZEFFECTS. SZ acknowledges support from the Academy of Finland FiDiPro progam, project 13282993. 
   
\newpage
\section*{Figure captions}   
   
\end{document}